\documentclass[letterpaper, 10 pt, conference]{ieeeconf}  % Comment this line out if you need a4paper

\IEEEoverridecommandlockouts                              % This command is only needed if 
\usepackage{graphics} % for pdf, bitmapped graphics files
\usepackage{epsfig} % for postscript graphics files
\usepackage{mathptmx} % assumes new font selection scheme installed
\usepackage{times} % assumes new font selection scheme installed
\usepackage{amsmath} % assumes amsmath package installed
\usepackage{amssymb}  % assumes amsmath package installed
\usepackage{mathtools}
\usepackage[flushleft]{threeparttable}
\usepackage{booktabs}
\usepackage{color}
\usepackage{float}

\title{\bf \large{Automatic Tracking of the Muscle Tendon Junction in Healthy and Impaired Subjects using Deep Learning*}}

\author{%
\parbox{15cm}{\centering%
\small{Christoph Leitner$^{1,2,\dagger}$, Robert Jarolim$^{3,\dagger}$, Andreas Konrad$^{2}$, Annika Kruse$^{2}$, Markus Tilp$^{2}$, J\"org Schr\"ottner$^{1}$ and Christian Baumgartner$^{1}$}\\
\tiny{© 2020 IEEE. Personal use of this material is permitted. Permission from IEEE must be obtained for all other uses, in any current or future media, including reprinting/republishing this material for advertising or promotional purposes, creating new collective works, for resale or redistribution to servers or lists, or reuse of any copyrighted component of this work in other works.}
}
\thanks{$^{1}$Institute of Health Care Engineering, Graz University of Technology, 8010 Graz, Austria. Corr.: {\tt\small christoph.leitner@tugraz.at}}%
\thanks{$^{2}$Institute of Sport Science, University of Graz, 8010 Graz, Austria.}%
\thanks{$^{3}$Institute of Physics, University of Graz, 8010 Graz, Austria}%
\thanks{$^{\dagger}$These authors contributed equally to this work.}%
\thanks{*This work was not supported by any organization.}% <-this % stops a space
}

\begin{document}

\maketitle
\thispagestyle{empty}
\pagestyle{empty}

%%%%%%%%%%%%%%%%%%%%%%%%%%%%%%%%%%%%%%%%%%%%%%%%%%%%%%%%%%%%%%%%%%%%
\begin{abstract}
Recording muscle tendon junction displacements during movement, allows separate investigation of the muscle and tendon behaviour, respectively. In order to provide a fully-automatic tracking method, we employ a deep learning approach to detect the position of the muscle tendon junction in ultrasound images. We utilize the attention mechanism to enable the network to focus on relevant regions and to obtain a better interpretation of the results. Our data set consists of a large cohort of 79 healthy subjects and 28 subjects with movement limitations performing passive full range of motion and maximum contraction movements. Our trained network shows robust detection of the muscle tendon junction on a diverse data set of varying quality with a mean absolute error of 2.55 $\pm$ 1 mm. We show that our approach can be applied for various subjects and can be operated in real-time. The complete software package is available for open-source use.
\newline
\\
\indent \textit{Clinical relevance}-— Muscle tendon dynamics during locomotion differ between healthy subjects and patients with e.g. cerebral palsy. Moreover, investigations of impaired individuals can be methodically difficult and image qualities vary due to transducer placement and movement tasks. We propose a reliable and time efficient method to track the muscle tendon junction and support clinical biomechanists in gait analysis.
\end{abstract}
%
%
%%%%%%%%%%%%%%%%%%%%%%%%%%%%%%%%%%%%%%%%%%%%%%%%%%%%%%%%%%%%%%%%%%%%%%%%%%%
%%%%%%%%%%%%%%%%%%%%%%%%%%%%%%%%%%%%%%%%%%%%%%%%%%%%%%%%%%%%%%%%%%%%%%%%%%%
\section{INTRODUCTION}
\label{sect:introduction}
Ultrasound (US) scanners enable real time views on muscles and tendons during human movement. Hence, the recording of local and distinct landmarks eg. the muscle tendon junction (MTJ) is used to investigate internal dynamics. Estimation methods and recorded MTJ movements allow the identification of individual contributions of muscles and tendons to the behaviour of whole muscle tendon units \cite{j:Leitner2020}, respectively. This is a well-established, in-vivo approach to investigate muscle and tendon properties in healthy \cite{j:Dick2016} and diseased subjects \cite{j:Barber2017}.

Manual tracking of the MTJ in large cohorts is time consuming. Fully and semi-automated tracker based on optical-flow and block-matching algorithms have been proposed. As shown on musculoskeletal investigations these approaches are prone to errors \cite{j:Cronin2011} when large displacements between imaging frames occur eg. in maximum voluntary contractions (MVCs). These large shifts in US image sequences arise mostly due to low frame rate recordings of fast tissue velocities \cite{j:Leitner2019}.

Deep learning methods have shown promising results in various fields \cite{lecun2015deep}. Especially in image processing, deep learning has achieved new state-of-the-art results with the use of convolutional neural networks (CNN). In most cases the interpretability of actions taken within the network is limited and even minor changes to the input can cause strong changes in the prediction \cite{adversarial_examples}. This delimits the credibility of the obtained results, nevertheless CNNs for classification tasks have shown the ability to localize features in the image \cite{discriminative_localization}. A recent advance in CNNs is the ability to focus on relevant regions in an image with the use of the attention mechanism. This method provides an increased stability against adversarial attacks and enables a direct localisation of the relevant regions \cite{Attention}.

In this work we show the use of CNNs to detect and track the MTJ, whereby we employ ResNet50 \cite{ResNet} and VGG16 \cite{VGG} architectures to solve this task. We demonstrate that valid results can be achieved using a deep learning approach. However, diverse tissue compositions of healthy and impaired subjects, the investigated movement task and the placement of the US probe can affect US image qualities. Therefore, we decided to further extend the VGG16 architecture with an attention mechanism \cite{Attention} to evaluate if the network predicts basing on relevant features and ignores errors in the data.

The subsequent study is divided into three sections. In Section \ref{sect:methods} we discuss the data set generation and data preprocessing. Moreover, we present adapted model architectures and employed training processes. In Section \ref{sect:results} we show the quantitative evaluation of our research. We conclude the study in Section \ref{sect:discussion} with a discussion of the results and a prospect on future applications.

Codes and pretrained models as well as high resolution images presented in this paper are publicly available via: \textcolor{blue}{https://github.com/luuleitner/deepMTJ}
%
%%%%%%%%%%%%%%%%%%%%%%%%%%%%%%%%%%%%%%%%%% MODEL ARCHITECTURE FIGURE
\begin{figure*}[thpb]
  \centering
  \framebox{\parbox{14.0cm}{
  \centering
  \includegraphics[scale=0.5]{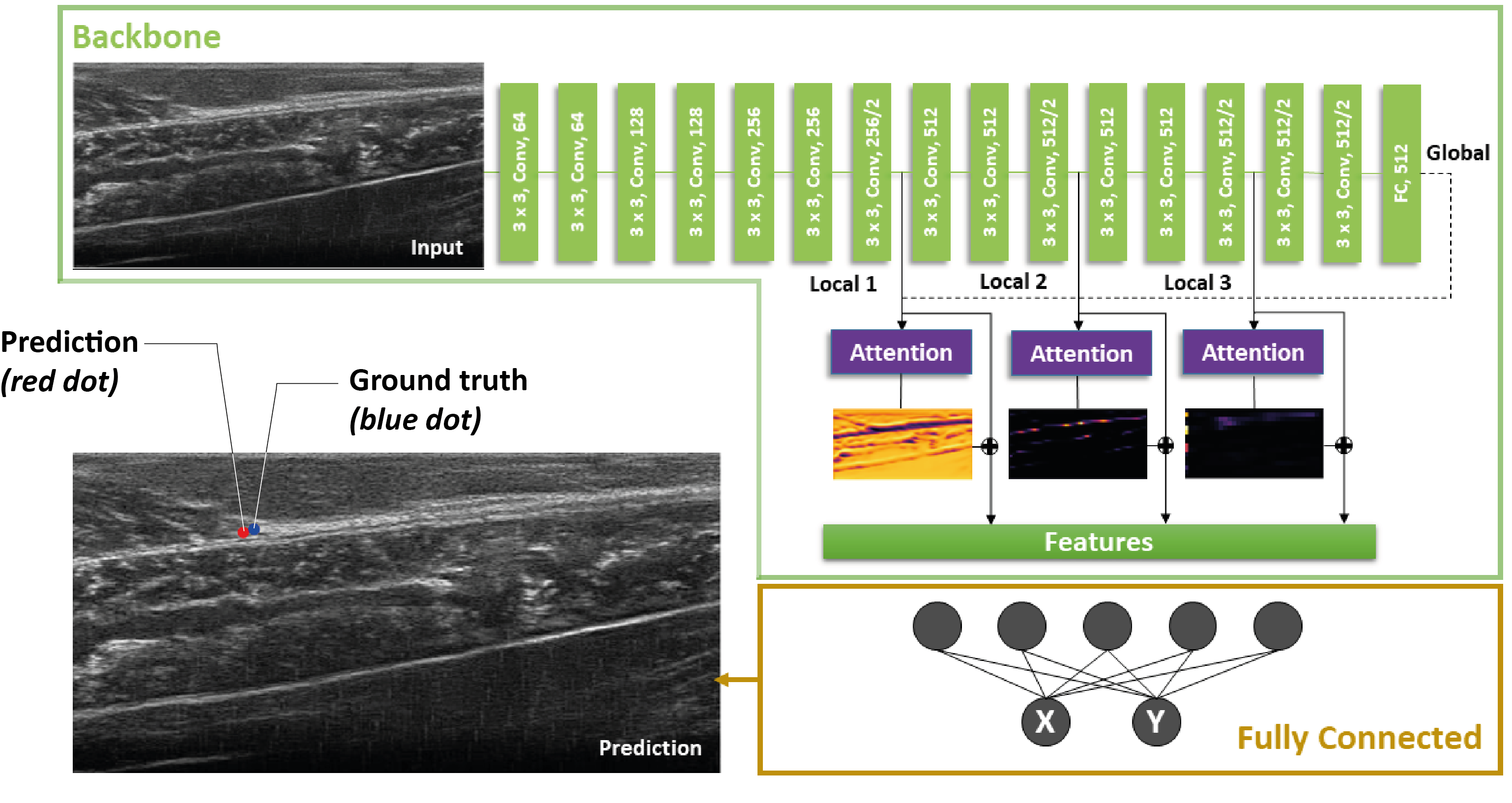}}} 
  \caption{Overview of the model architecture with the VGG-Attention backbone. Three attention maps are created from the correlation of the extracted local features with the global feature (see \cite{Attention} for comparison). MTJ prediction \textit{(red dot)} and ground truth \textit{(blue dot).}}
  \label{fig:model}
\end{figure*}
\begin{table}[h]
\caption{Our data set was generated in six studies.}
\label{tab:studies}
\begin{center}
%\hspace*{-1.0cm} 
\renewcommand{\arraystretch}{1.20}
\begin{tabular}{cccc}
\toprule
\textbf{Studies}       & \textbf{Movements} & \textbf{Subjects} & \textbf{H/I$^\dagger$} \\ 
\midrule
Konrad et al. \cite{j:Konrad2014:1,j:Konrad2015,j:Konrad2014:2} & MVC, ROM  & 66       & 66 / 0     \\
Kruse et al. \cite{j:Kruse2017,j:Kruse2018,j:Kruse2019}  & MVC, ROM  & 41       & 13 / 28            \\ 
\bottomrule
\end{tabular}
 \begin{tablenotes}
    \item MVC ... maximum voluntary contraction, ROM ... passive range of motion, $\dagger$ Healthy/Impaired 
    \end{tablenotes}
\end{center}
\end{table}
%
%
%
%%%%%%%%%%%%%%%%%%%%%%%%%%%%%%%%%%%%%%%%%%%%%%%%%%%%%%%%%%%%%%%%%%%%%%%%%%%
%%%%%%%%%%%%%%%%%%%%%%%%%%%%%%%%%%%%%%%%%%%%%%%%%%%%%%%%%%%%%%%%%%%%%%%%%%%
\section{METHODS} 
\label{sect:methods}
%
%
%123456789012345678901234567890123456789012345678901234567890
\subsection{Data Set}
Training data was collected from 6 different studies (Table \ref{tab:studies}) conducted at the Institute of Sport Science at the University of Graz. Studies investigated MVC and passive range of motion (ROM) measurments on 79 healthy subjects and 28 patients with cerebral palsy. Research was approved by the ethics committees of the Medical University of Graz (21-362) and the University of Graz (39/77/63). 

The fully anonymous data set holds 1106 videos with a length of 15 seconds/video. Sequences were recorded at frame rates of 25 Hz. For model training and testing we first randomly selected 12 subjects and then completely separated the related 127 videos into a test data set. The remaining number of 979 videos saw a 90\% training -- 10\% validation set split (refer to Table \ref{tab:dataset} for absolute numbers).
%
%%%%%%%%%%%%%%%%%%%%%%%%%%%%%%%%%%%%%%%%%% DATA SPECIFICATIONS TABLE
\begin{table}[h]
\caption{Dataset Specifications}
\label{tab:dataset}
\begin{center}
\renewcommand{\arraystretch}{1.2}
\hspace*{-1.0cm} 
\setlength{\tabcolsep}{3pt} % Default value: 6pt
\begin{tabular}{cccc}
\toprule
\textbf{Specification}      & \textbf{Value}\\ 
\midrule
Scan mode    & B-mode   \\ 
Center frequency & 7 MHz    \\ 
Framerate       & 25 Hz \\
Image size      & 128x64 px \\
Training set     &7.200 frames  \\
Validation set  &800 frames    \\
Test set         & 1147 frames  \\
\bottomrule
\end{tabular}
\end{center}
\end{table}

The data set was annotated by an experienced investigator using a python coded annotation tool (part of provided software package) at a sampling frequency of 10 frames per video to avoid strong temporal correlations. The position of the MTJ was identified as the most distal insertion of the muscle into the free tendon (Fig. \ref{fig:model}). The annotation of the data set was conducted within 10 days. To keep the annotation bundles small and guarantee reproducibility, the data set (training, validation and test) was split into small batches and annotated during 7 days. An annotation review of the complete set was conducted 2 days after the completion of the primary annotation round. The annotated images were cropped to an image size of 512x256 pixels and afterwards rescaled to 128x64 pixels. The images are normalized to an interval between -1 and 1. The coordinates of the MTJ are scaled between 0 and 1 for each axis. The results of this annotation process are further referred to as ground truth.
%
%%%%%%%%%%%%%%%%%%%%%%%%%%%%%%%%%%%%%%%%%% OTHER TRACKING ALGORITHMS TABLE
\begin{table*}[thpb]
\caption{Comparison of identified research on automatic tracking of the muscle tendon junction in 2D B-Mode ultrasound images}
\centering
\label{tab:algorithms}
\begin{center}
%\hspace*{-1.0cm}
\setlength{\tabcolsep}{3pt} % Default value: 6pt
\renewcommand{\arraystretch}{1.5}
\begin{tabular}{cccccccc}
\toprule
\textbf{Study}  & \textbf{Subjects} & \textbf{H / I$^\dagger$}   & \textbf{Tested Movements}  & \textbf{Method}  &\textbf{Mean Error [mm]}    &\textbf{Calculation time (200 frames)  [s] }\\ 
\midrule
*Cenni (2019) \cite{j:Cenni2019} &32 & 20 / 12   &stretching, walking    &optical flow &$1.7\pm1.9 - 20.5\pm5.6$  &11-14\\
Zhou (2018) \cite{j:Zhou2018}   &10   &10 / 0   &ROM  &optical flow  &--    &400-600\\ 
Lee (2008) \cite{j:Lee2008} &5 &5 / 0 &ROM &optical flow  &--$^\ddagger$    &--$^\ddagger$  \\  
\textbf{*This work:} & \textbf{107}   & \textbf{79 / 28}  &\textbf{MVC, ROM}  &\textbf{VGG-Attention-3} &2.55 $\pm$ 1.00 &1.5\\
\bottomrule
\end{tabular}
 \begin{tablenotes}
    \item * open source, MVC ... maximum voluntary contraction, ROM ... passive range of motion
    \item $\dagger$ Healthy / Impaired
    \item $\ddagger$ Lee et al. \cite{j:Lee2008} listed an RMSE of $5.9\pm1.8\%-7.9\pm2.4\%$ between manual and automatic tracking of their data set on in-vivo experiments. Cenni et al. \cite{j:Cenni2019} published an improved open source software library based on Lee's algorithm delivering better results.
    \end{tablenotes}
\end{center}
\end{table*}
%
%
%123456789012345678901234567890123456789012345678901234567890
\subsection{Model Architecture}

To track the MTJ we restrict the model to a single point detection. Hence, the network predicts the x- and y-coordinates of the MTJ based on the full image input. We evaluated three different model architectures (Table \ref{tab:results}). Our proposed models consist of a backbone which follows the original architectures. The classification layers are replaced by two sigmoid output units, which resemble the pixel position of the MTJ in the image. 

The general approach for each architecture is to transform the input per convolutional layer, hereby the spatial dimensions are reduced while the depth of the network is increased. The ResNet50 approach adds skip connections between the convolutional blocks to overcome the vanishing gradient problem and leads to improved results with reduced training time \cite{Attention}. 

For the detection of the MTJ only a small fraction of the image is relevant. To direct the focus on relevant regions in the input image we follow the approach by Jetlay et al. \cite{Attention} and introduce attention modules into our VGG16 architecture. The proposed model extracts up to three local feature vectors. They are then correlated via a dot product with the global extracted feature vector. Attention maps are calculated using a softmax function on the dot product. Furthermore, these maps are applied on the extracted local feature vectors. This enables the network to suppress irrelevant information while keeping the focus on important and mature features in the image (Fig. \ref{fig:model}). We refer to VGG-Attention-2 and VGG-Attention-3 as the VGG16 architecture with 2 and 3 attention modules, respectively.
%
%123456789012345678901234567890123456789012345678901234567890
\subsection{Training}
\label{sect:training}
We optimize each of the proposed models for the squared euclidean distance between prediction and ground truth, in order to penalize large deviations stronger. Since the x- and y-coordinates of the labels are scaled between [0,1], the loss is accordingly weighted to the image dimensions as given by Eq. \ref{eq:loss} with X and Y as ground truth and $\hat{X}$ and $\hat{Y}$ as predictions.
\begin{equation}
\label{eq:loss}
    loss = \frac{1}{5}\left[4(X - \hat{X})^2 + (Y - \hat{Y})^2\right]
\end{equation}
To train the ResNet50 and VGG16 neural networks we use the Adam optimizer \cite{adam_optimizer} with an initial learning rate of $\alpha=0.001$. For the architectures with the attention modules, we follow the approach by Jetley et al. \cite{Attention} and use stochastic gradient descend with a learning rate of $\alpha=0.01$, momentum of 0.9 and a l2 weight decay of 0.0005.
We adaptivly reduce the learning rates step-wise, by scaling them with a factor of 0.5 every 25 epochs. We employ a batch size of 32 and train for 300 epochs until convergence is reached.
%
%123456789012345678901234567890123456789012345678901234567890
\subsection{Quantitative Evaluation Metrics}
To evaluate performance and robustness of our trained networks, we discuss four different metrics.

To assess the tracking performances between our benchmark models we calculate the root mean square error (RMSE) (Eq. \ref{eq:rmse}) of the predicted MTJ ($\hat{\theta}$) and ground truth labels ($\theta$) of the test set.
\begin{equation}
\label{eq:rmse}
    RMSE = \sqrt{\dfrac{\sum_{i=1}^{n} (\theta_i-\hat{\theta_i})^2}{n}}
\end{equation}

\begin{figure}[thpb]
    \centering
    \framebox{\parbox{7.5cm}{
    \centering
    \includegraphics[scale=0.398]{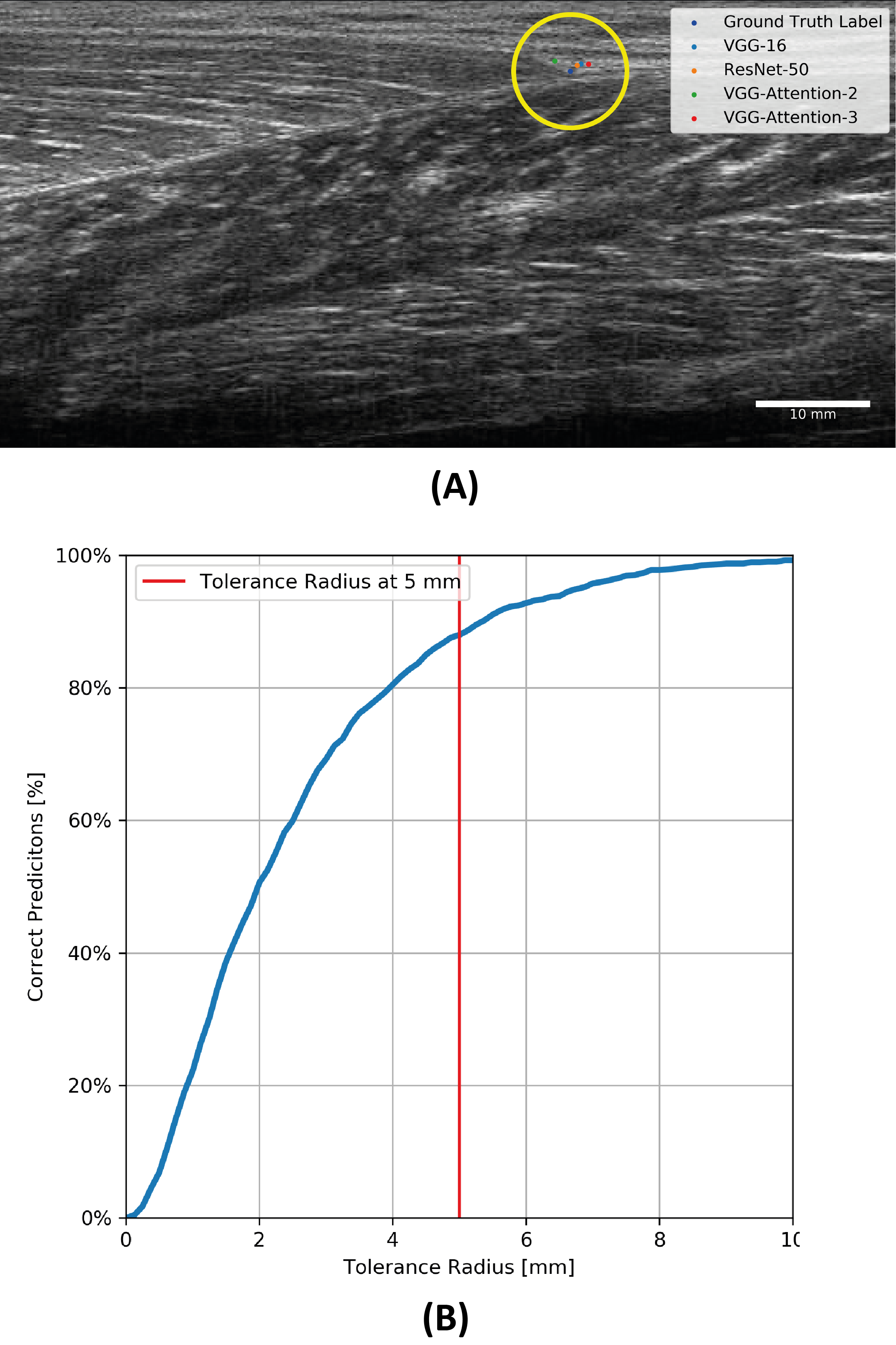}
}}
    \caption{(A) Example of the MTJ predictions (VGG16 \textit{(light blue dot)}, ResNet50 \textit{(orange dot)}, VGG-Attention-2 \textit{(green dot)}, VGG-Attention-3 \textit{(red dot)}) and the ground truth label \textit{(blue dot)}. The 5 mm tolerance radius \textit{(yellow circle)} indicates the classification limit as valid prediction. (B) Percentage of correct MTJ predictions over increasing distances from the labeled ground truth, based on the results of the VGG-Attention-3 model. The red line marks the valid predictions at a tolerance radius of 5mm.}
    \label{fig:tolerance}
\end{figure}

\begin{figure*}[thpb]
   \centering
    \framebox{\centering\parbox{17.8cm}{
    \includegraphics[scale=0.98]{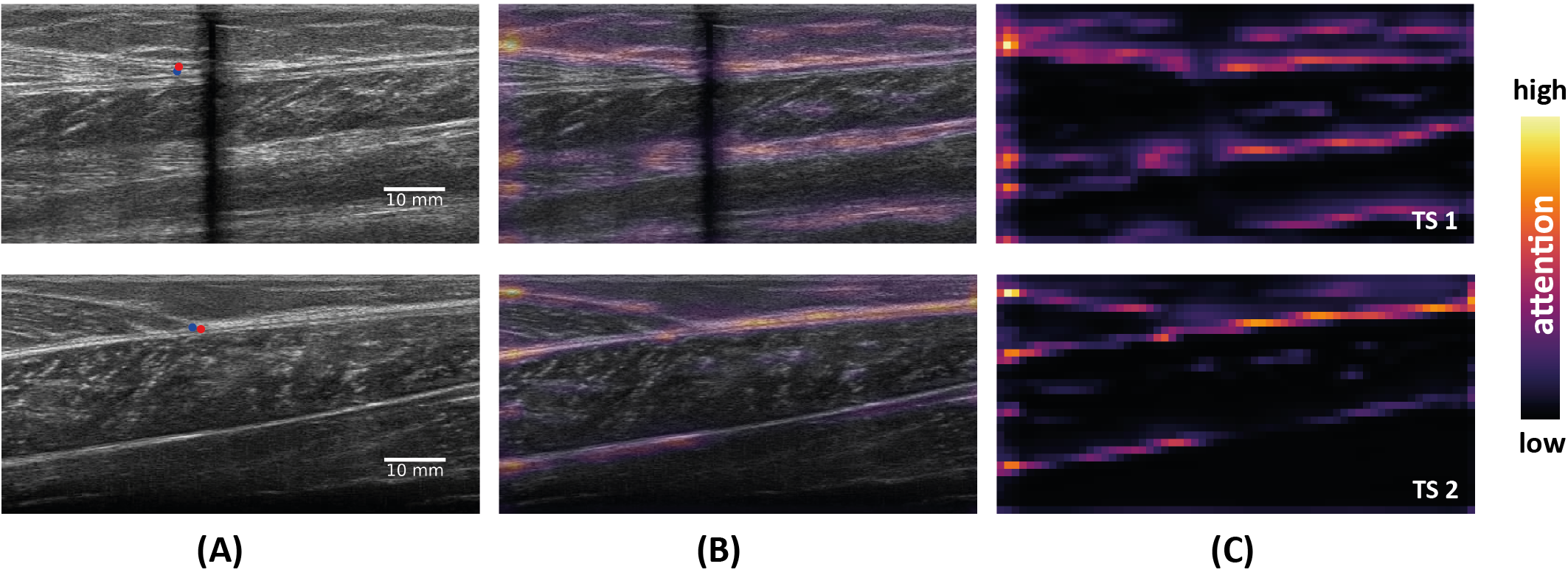}
}}
    \caption{Results of test sample 1 (TS 1) in the upper row and test sample 2 (TS 2) in the lower row calculated with the VGG-Attention-3 model: (A) MTJ prediction \textit{(red dot)} and ground truth \textit{(blue dot)} (B) Overlay of attention map with original image (C) Combined attention map. TS 1 shows that the attention module neglects distortions in the image.}
    \label{fig:attention2}
\end{figure*}

Furthermore, we calculate the deviation between the predicted MTJ and the ground truth labels in terms of the mean-absolute-error (MAE) in mm. We introduce a tolerance radius of 5 mm centered around the ground truth label to classify the predctions of the MTJ as either valid or invalid (Fig. \ref{fig:tolerance}).
Moreover, we investigate the extracted attention maps and assess their correspondence with the MTJ prediction. For visualization of the relevant features the last two attention maps are re-scaled and  pixel-wise multiplied to obtain a combined attention map (Fig. \ref{fig:attention2}).
%
%%%%%%%%%%%%%%%%%%%%%%%%%%%%%%%%%%%%%%%%%% TOLERANCE RESULTS FIGURE
%
%
%%%%%%%%%%%%%%%%%%%%%%%%%%%%%%%%%%%%%%%%%% ATTENTIONMAP RESULTS FIGURE
%
%
%
%%%%%%%%%%%%%%%%%%%%%%%%%%%%%%%%%%%%%%%%%%%%%%%%%%%%%%%%%%%%%%%%%%%%%%%%%%%
%%%%%%%%%%%%%%%%%%%%%%%%%%%%%%%%%%%%%%%%%%%%%%%%%%%%%%%%%%%%%%%%%%%%%%%%%%%
\section{RESULTS} 
\label{sect:results}
After model training as prescribed in section \ref{sect:training} we evaluated them on a previously separated test set where we explicitly excluded subjects from the training set. An overview on model performances is presented in Table \ref{tab:results}.

Furthermore, Fig. \ref{fig:tolerance}B shows the percentage of correct detections in dependence of the growing tolerance radius.

Extracted attention maps for a clean test sample (TS 2, lower row) and a distorted image (TS 1, upper row) calculated with the VGG-Attention-3 architecture are shown in Fig. \ref{fig:attention2}. 
%
%%%%%%%%%%%%%%%%%%%%%%%%%%%%%%%%%%%%%%%%%% RESULTS TABLE
\begin{table}[h]
\caption{Performance Overview of Models on Test Set Data}
\label{tab:results}
\begin{center}
\hspace*{-1.55cm}
\renewcommand{\arraystretch}{1.25}
\begin{tabular}{cccc}
\toprule
\textbf{Model}  & \textbf{MAE [mm]} & \textbf{Valid [\%]*} &\textbf{RMSE [\%]}\\ 
\midrule
VGG16 \cite{VGG}  & 2.66 $\pm$ 1.24 & 86.14  & 4.46   \\
ResNet50 \cite{ResNet}  & 4.37 $\pm$ 1.63   & 70.62  & 7.13   \\
VGG-Attention-2 \cite{Attention}  & 4.98 $\pm$ 1.64    & 64.25  &  7.48  \\
\textbf{VGG-Attention-3 \cite{Attention}}  & \textbf{2.55 $\pm$ 1.00}    & \textbf{88.32}  &  \textbf{3.75}  \\
\bottomrule
\end{tabular}
\begin{tablenotes}
    \item * Tolerance radius 5mm
    \end{tablenotes}
\end{center}
\end{table}
%
%
%
%%%%%%%%%%%%%%%%%%%%%%%%%%%%%%%%%%%%%%%%%%%%%%%%%%%%%%%%%%%%%%%%%%%%%%%%%%%
%%%%%%%%%%%%%%%%%%%%%%%%%%%%%%%%%%%%%%%%%%%%%%%%%%%%%%%%%%%%%%%%%%%%%%%%%%%
\section{DISCUSSION \& CONCLUSIONS} 
\label{sect:discussion}
Our performance evaluations are based on an independent and completely separated test set. Subjects performing MVCs or ROMs are either part of the training or the test set. We demonstrate that the proposed VGG-Attention-3 architecture with a RMSE of 3.75\% outperforms other model architectures (Table \ref{tab:results}). Furthermore, the MAE between predictions and ground truth labels is 2.55 mm for the VGG-Attention-3 model and provides a stronger performance than previously proposed tracking algorithms (Table \ref{tab:algorithms}). 

Fig. \ref{fig:attention2} indicates that the network identifies the reflective and distinct tendinous Y-shape in ultrasound images to predict the MTJ-position. Image distortions due to probe placement or movement tasks can lead to poor image qualities and errors in the prediction of automated extraction methods. The extracted attention maps show that the network correctly ignores distorted regions (Fig. \ref{fig:attention2}). Moreover, this method can operate in real-time, which makes MTJ predictions more than 7x faster than previously developed tracking methods based on optical flow or block matching algorithms (Table \ref{tab:algorithms}). 

Our approach has shown the capability to track muscle tendon junctions in ultrasound images in a precise and time efficient way. Our data set is characterized by a large number (n=107) of diverse (healthy and impaired) subjects. It includes full range of motion and maximum contractions. The total volume of our data set is larger than in previous approaches (training 6400 frames/validation 1600 frames/test 1147 frames). Due to the clear separation of individual subjects into a training and test set, we show that our approach is capable of tracking the MTJ on previously unseen subjects.

We note that the generalization to different instruments (eg. ultrasound systems) has not been analysed within this study and will be part of our future investigations. Our codes and pretrained models are publicly available, to foster further use of our methods in clinical as well as research applications.

%%%%%%%%%%%%%%%%%%%%%%%%%%%%%%%%%%%%%%%%%%%%%%%%%%%%%%%%%%%%%%%%%%%%%%%%%%%%%%%%

%%%%%%%%%%%%%%%%%%%%%%%%%%%%%%%%%%%%%%%%%%%%%%%%%%%%%%%%%%%%%%%%%%%%%%%%%%%%%%%%
%
%
%%%%%%%%%%%%%%%%%%%%%%%%%%%%%%%%%%%%%%%%%%%%%%%%%%%%%%%%%%%%%%%%%%%%%%%%%%%
%%%%%%%%%%%%%%%%%%%%%%%%%%%%%%%%%%%%%%%%%%%%%%%%%%%%%%%%%%%%%%%%%%%%%%%%%%%
%\section*{APPENDIX}
%A systematic search identified studies that published automatic tracking algorithms of the muscle tendon junction in images or videos. The literature review included the following databases: Web of Science, SCOPUS and Medline. The database search revealed 64 items, and 2 other sources were added. After title and abstract screening, 7 full-text articles were considered eligible. Four were excluded as they did not meet the criteria. In conclusion 3 studies (Table 1) were identified to meet the inclusion critera (automatic, muscle tendon junction, tracking).
%
%
%%%%%%%%%%%%%%%%%%%%%%%%%%%%%%%%%%%%%%%%%%%%%%%%%%%%%%%%%%%%%%%%%%%%%%%%%%%
%%%%%%%%%%%%%%%%%%%%%%%%%%%%%%%%%%%%%%%%%%%%%%%%%%%%%%%%%%%%%%%%%%%%%%%%%%%
\section*{ACKNOWLEDGMENT}
C. Leitner would like to acknowledge Martin Sust (University of Graz) for the support of his research and Luca Benini (ETH Zurich and Universit\`{a} di Bolognia) for his insightful feedback.
%
%
%%%%%%%%%%%%%%%%%%%%%%%%%%%%%%%%%%%%%%%%%%%%%%%%%%%%%%%%%%%%%%%%%%%%%%%%%%%
%%%%%%%%%%%%%%%%%%%%%%%%%%%%%%%%%%%%%%%%%%%%%%%%%%%%%%%%%%%%%%%%%%%%%%%%%%%
\bibliographystyle{IEEEtran}
\bibliography{references.bib}

\begin{thebibliography}{10}
\providecommand{\url}[1]{#1}
\csname url@samestyle\endcsname
\providecommand{\newblock}{\relax}
\providecommand{\bibinfo}[2]{#2}
\providecommand{\BIBentrySTDinterwordspacing}{\spaceskip=0pt\relax}
\providecommand{\BIBentryALTinterwordstretchfactor}{4}
\providecommand{\BIBentryALTinterwordspacing}{\spaceskip=\fontdimen2\font plus
\BIBentryALTinterwordstretchfactor\fontdimen3\font minus
  \fontdimen4\font\relax}
\providecommand{\BIBforeignlanguage}[2]{{%
\expandafter\ifx\csname l@#1\endcsname\relax
\typeout{** WARNING: IEEEtran.bst: No hyphenation pattern has been}%
\typeout{** loaded for the language `#1'. Using the pattern for}%
\typeout{** the default language instead.}%
\else
\language=\csname l@#1\endcsname
\fi
#2}}
\providecommand{\BIBdecl}{\relax}
\BIBdecl

\bibitem{j:Leitner2020}
\BIBentryALTinterwordspacing
C.~Leitner, C.~Baumgartner, C.~Peham, and M.~Tilp, ``{Ultrasound in Locomotion
  Research -- The Quest for Wider Views},'' \emph{arXive}, 1 2020. [Online].
  Available: \url{http://arxiv.org/abs/2001.06718}
\BIBentrySTDinterwordspacing

\bibitem{j:Dick2016}
T.~J. M.~T. Dick, A.~A.~S. Arnold, and J.~M. Wakeling,
  ``\BIBforeignlanguage{eng}{{Quantifying Achilles tendon force in vivo from
  ultrasound images.}}'' \emph{\BIBforeignlanguage{eng}{Journal of
  biomechanics}}, vol.~49, no.~14, pp. 3200--3207, 10 2016.

\bibitem{j:Barber2017}
L.~Barber, C.~Carty, L.~Modenese, J.~Walsh, R.~Boyd, and G.~Lichtwark,
  ``{Medial gastrocnemius and soleus muscle-tendon unit, fascicle, and tendon
  interaction during walking in children with cerebral palsy},''
  \emph{Developmental Medicine {\&} Child Neurology}, vol.~59, no.~8, pp.
  843--851, 8 2017.

\bibitem{j:Cronin2011}
N.~J. Cronin, C.~P. Carty, R.~S. Barrett, and G.~Lichtwark, ``{Automatic
  tracking of medial gastrocnemius fascicle length during human locomotion},''
  \emph{Journal of Applied Physiology}, vol. 111, no.~5, pp. 1491--1496, 11
  2011.

\bibitem{j:Leitner2019}
C.~Leitner, P.~A. Hager, H.~Penasso, M.~Tilp, L.~Benini, C.~Peham, and
  C.~Baumgartner, ``{Ultrasound as a Tool to Study Muscle–Tendon Functions
  during Locomotion: A Systematic Review of Applications},'' \emph{Sensors},
  vol.~19, no.~19, p. 4316, 10 2019.

\bibitem{lecun2015deep}
Y.~LeCun, Y.~Bengio, and G.~Hinton, ``Deep learning,'' \emph{nature}, vol. 521,
  no. 7553, p. 436, 2015.

\bibitem{adversarial_examples}
I.~J. Goodfellow, J.~Shlens, and C.~Szegedy, ``Explaining and harnessing
  adversarial examples,'' \emph{arXiv preprint arXiv:1412.6572}, 2014.

\bibitem{discriminative_localization}
B.~Zhou, A.~Khosla, A.~Lapedriza, A.~Oliva, and A.~Torralba, ``Learning deep
  features for discriminative localization,'' in \emph{Proceedings of the IEEE
  conference on computer vision and pattern recognition}, 2016, pp. 2921--2929.

\bibitem{Attention}
S.~Jetley, N.~A. Lord, N.~Lee, and P.~H. Torr, ``Learn to pay attention,''
  \emph{arXiv preprint arXiv:1804.02391}, 2018.

\bibitem{ResNet}
K.~He, X.~Zhang, S.~Ren, and J.~Sun, ``Deep residual learning for image
  recognition,'' in \emph{Proceedings of the IEEE conference on computer vision
  and pattern recognition}, 2016, pp. 770--778.

\bibitem{VGG}
K.~Simonyan and A.~Zisserman, ``Very deep convolutional networks for
  large-scale image recognition,'' \emph{arXiv preprint arXiv:1409.1556}, 2014.

\bibitem{j:Konrad2014:1}
A.~Konrad and M.~Tilp, ``{Increased range of motion after static stretching is
  not due to changes in muscle and tendon structures},'' \emph{Clinical
  Biomechanics}, vol.~29, no.~6, pp. 636--642, 2014.

\bibitem{j:Konrad2015}
A.~Konrad, M.~Gad, and M.~Tilp, ``{Effect of PNF stretching training on the
  properties of human muscle and tendon structures},'' \emph{Scandinavian
  Journal of Medicine and Science in Sports}, vol.~25, no.~3, pp. 346--355, 6
  2015.

\bibitem{j:Konrad2014:2}
A.~Konrad and M.~Tilp, ``{Effects of ballistic streching training on the
  properties of human muscle and tendon structures},'' \emph{Journal of Applied
  Physiology}, vol. 117, no.~1, pp. 29--35, 7 2014.

\bibitem{j:Kruse2017}
A.~Kruse, C.~Schranz, M.~Svehlik, and M.~Tilp, ``{Mechanical muscle and tendon
  properties of the plantar flexors are altered even in highly functional
  children with spastic cerebral palsy},'' \emph{Clinical Biomechanics},
  vol.~50, pp. 139--144, 12 2017.

\bibitem{j:Kruse2018}
A.~Kruse, C.~Schranz, M.~Tilp, and M.~Svehlik, \emph{{Muscle and tendon
  morphology alterations in children and adolescents with mild forms of spastic
  cerebral palsy}}.\hskip 1em plus 0.5em minus 0.4em\relax BMC Pediatrics, 12
  2018, vol.~18.

\bibitem{j:Kruse2019}
A.~Kruse, C.~Schranz, M.~Svehlik, and M.~Tilp, ``{The effect of functional
  home-based strength training programs on the mechano-morphological properties
  of the plantar flexor muscle-tendon unit in children with spastic cerebral
  palsy},'' \emph{Pediatric Exercise Science}, vol.~31, no.~1, pp. 67--75, 2
  2019.

\bibitem{j:Cenni2019}
F.~Cenni, L.~Bar-On, D.~Monari, S.-H. Schless, B.~Kalkman, E.~Aertbeli{\"{e}}n,
  K.~Desloovere, and H.~Bruyninckx, ``{Semi-automatic methods for tracking the
  medial gastrocnemius muscle–tendon junction using ultrasound: a validation
  study},'' \emph{Experimental Physiology}, vol. 105, no.~1, pp. 120--131,
  2020.

\bibitem{j:Zhou2018}
G.-Q. Zhou, Y.~Zhang, R.-L. Wang, P.~Zhou, Y.-P. Zheng, O.~Tarassova, A.~Arndt,
  and Q.~Chen, ``{Automatic Myotendinous Junction Tracking in Ultrasound Images
  with Phase-Based Segmentation},'' \emph{BioMed Research International}, vol.
  2018, pp. 1--12, 2018.

\bibitem{j:Lee2008}
S.~Lee, G.~Lewis, and S.~Piazza, ``{An algorithm for automated analysis of
  ultrasound images to measure tendon excursion in vivo},'' \emph{Journal of
  Applied Biomechanics}, vol.~24, no.~1, pp. 75--82, 2008.

\bibitem{adam_optimizer}
D.~P. Kingma and J.~Ba, ``Adam: A method for stochastic optimization,''
  \emph{arXiv preprint arXiv:1412.6980}, 2014.

\end{thebibliography}
\end{document}